\documentclass[aps,pra,onecolumn,superscriptaddress,showpacs,10pt,amsmath,footinbib,preprint]{revtex4-1}
\usepackage{amsthm,graphicx,color}
\usepackage{wasysym,mathtools,placeins}
\usepackage{floatrow,soul,enumitem}
\usepackage{hyperref}

\newcommand{\Lz}{\left|L_z\right|}
\providecommand{\gtsimeq}{\raisebox{-0.6ex}{$\,\stackrel
	{\raisebox{-.2ex}{$\textstyle >$}}{\sim}\,$}}
\providecommand{\ltsimeq}{\raisebox{-0.6ex}{$\,\stackrel
	{\raisebox{-.2ex}{$\textstyle <$}}{\sim}\,$}}

\begin{document}
\title{From vortices to solitonic vortices in trapped atomic Bose-Einstein condensates}

\author{M. C. Tsatsos}
\thanks{Corresponding author}\email{marios@ifsc.usp.br}
\affiliation{S\~ao Carlos Institute of Physics, University of S\~ao Paulo, P.O, Box 369, 13560-970 S\~ao Carlos, SP, Brazil}
\author{M. J. Edmonds}
\affiliation{Joint Quantum Centre (JQC) Durham-Newcastle, School of Mathematics and Statistics, Newcastle University, Newcastle upon Tyne NE1 7RU, England, United Kingdom}
\author{N. G. Parker}
\affiliation{Joint Quantum Centre (JQC) Durham-Newcastle, School of Mathematics and Statistics, Newcastle University, Newcastle upon Tyne NE1 7RU, England, United Kingdom}

\date{\today}

\begin{abstract}
Motivated by recent experiments we study theoretically the dynamics of vortices in the crossover from two to one-dimension in atomic condensates in elongated traps.  We explore the transition from the dynamics of a vortex to that of a dark soliton as the one-dimensional limit is approached, mapping this transition out as a function of the key system parameters. Moreover, we probe this transition dynamically through the hysteresis under time-dependent deformation of the trap at the dimensionality crossover.  When the solitonic regime is probed during the hysteresis, significant angular momentum is lost from the system but, remarkably, the vortex can re-emerge.
\end{abstract}

\maketitle

{\it Introduction.} Atomic Bose-Einstein condensates (BECs) provide rich insight into superfluidity, buoyed by their purity and immense ability to control and image the coherent matter wave \cite{Bloch2008}. Of particular interest are coherent macroscopic excitations in the form of quantized vortices \cite{Fetter2009,Anderson2010} and dark solitons \cite{Frantzeskakis2010}. Quantized vortices represent defects in the quantum mechanical phase about which the superfluid flows with quantized circulation and appear point-like in 2D and as vortex lines or rings in 3D. Early landmark demonstrations of single vortices \cite{Matthews1999,Rosenbusch2002}, vortex arrays and lattices \cite{Madison2000,Aboshaeer2001,Hodby2001} and vortex rings \cite{Anderson2001} have been supplemented more recently by the deterministic generation of vortex dipoles \cite{Neely2010,Kwon2015}, real-time observation of vortex dynamics \cite{Freilich2010} and turbulent states of disordered vortices \cite{Henn2009, Neely2013,Kwon2014}. 
Meanwhile, atomic dark solitons are one-dimensional, non-dispersive matter waves characterized by a notch in the atomic density and a nontrivial phase slip \cite{Reinhardt1997,Frantzeskakis2010}. They are favored under repulsive {\it s}-wave atomic interactions, which gives rise to the required defocussing mean-field nonlinearity. Experiments have controllably generated dark solitons \cite{Burger1999,Denschlag2000,Dutton2001,Engels2007,Jo2007,Becker2008,Chang2008,Stellmer2008,Weller2008}, 
including long-lived solitons at ultralow temperatures in tightly-1D geometries \cite{Becker2008} and studied their oscillations, interactions and collisions \cite{Becker2008,Stellmer2008,Weller2008}.  

Dark solitons and vortices are formally distinct objects with differing dimensionalities and topological properties; vortices can only disappear at a boundary or by annihilating with an opposite-circulation vortex, while dark solitons have no such constraint. In a harmonic trap, a dark soliton tends to oscillate axially at a fixed proportion of the trap frequency  \cite{Fedichev1999,Muryshev1999,Busch2000,Huang2002} while a vortex precesses about the trap center at a frequency with a non-trivial dependence on its position and system parameters \cite{Svidzinsky2000a,Svidzinsky2000b,Lundh2000,Kim2004,Middelkamp2010}. Remarkably, however, dark solitons and vortices show many analogous behaviors -- underpinned by their common nature as phase defects -- such as their spontaneous creation under the Kibble-Zurek mechanism \cite{Lamporesi2013,Weiler2008,Middelkamp2011}, their emergence during the breakdown of superflow \cite{Frisch1992,Inouye2001,Neely2010,Kwon2014,Hakim1997,Engels2007}, their instability to acceleration \cite{Parker2003,Parker2004} and their interaction with phonons \cite{Allen2011,Parker2012}. 

The intimate connection between vortices and dark solitons is perhaps best revealed at the dimensionality crossover.  While dark solitons are dimensionally stable in quasi-1D geometries \cite{Theocharis2007}, 3D dark solitons are unstable to transverse perturbations; the nodal line undergoes the \emph{snake instability} (known from earlier studies in optics \cite{Kivshar1998}) and decays into one or more vortex rings (or vortex-antivortex pairs in 2D) \cite{Feder2000,Carr2000, Anderson2001,Dutton2001}. Close to the 1D boundary, hybrid dark soliton-vortex ring excitations have been observed \cite{Ginsberg2005}. Theoretical analysis of the possible solutions confirmed this behavior but also predicted the existence of solitonic vortex solutions \cite{Brand2002,Komineas2003}, that is, a single vortex confined to move along the long axis. This excitation is predicted to be favored when the transverse size is large enough to make the dark soliton unstable but not so large as to support vortex rings. In recent experiments these solitonic vortices have been reported in both Bose \cite{Becker2013,Donadello2014} and Fermi gases \cite{Ku2014}.

Motivated by these recent experiments we examine the crossover from vortices to solitonic vortices in trapped condensates. Based on numerical simulations of the 2D Gross-Pitaevskii (GP) equation, we investigate the propagation of the vortex/solitonic vortex in static traps with differing aspect ratios. We map out how the oscillation frequency of the excitation changes with the trap ratio; as the latter increases, this frequency saturates to that expected for a dark soliton, marking the onset of the solitonic vortex regime.  This occurs when the transverse harmonic oscillator length becomes roughly equal to twice the healing length (the characteristic size of the vortex).  Furthermore, we examine the dynamics in traps with time-dependent trap ratio, exploring the hysteresis across the vortex/solitonic vortex crossover. We find that observable deviations of the angular momentum from its initial value can occur if the vortex/solitonic vortex limit is crossed.

{\it Theoretical model.} We consider a weakly-interacting BEC at zero temperature, composed of atoms of mass $m$ and confined by a harmonic potential 
 $V(x,y,z)=\frac{1}{2}m(\omega_x^2 x^2+\omega_y^2 y^2 + \omega_{z}^{2}z^{2})$, where $\omega_{x,y,z}$ are the trap frequencies in the respective directions. The atomic interactions are modelled by the contact pseudopotential $g_0 \delta ({\rm r}-{\rm r}')$, where $g_0=4\pi\hbar^2a_{\rm s}/m$ and $a_{\rm s}$ is the atomic {\it s}-wave scattering length.  For simplicity we adopt a 2D model.  The trapping along $z$ is assumed sufficiently strong to render the condensate dynamics as quasi-two-dimensional \cite{Gorlitz2001}.  Then, the 2D condensate wave function $\psi(x,y,t)$ (normalized to the total particle number $N$) satisfies the effective 2D GP equation
\begin{equation}
i \hbar \frac{\partial \psi}{\partial t}=\left[-\frac{\hbar^2}{2m}\nabla^2+V(x,y,t)+\frac{g_0}{\sqrt{2\pi} l_z} |\psi|^2 \right]\psi,
\label{eq:gpe}
\end{equation}
where $l_z=\sqrt{\hbar/m \omega_z}$ is the harmonic oscillator length along $z$. 
The energy scale of the condensate is characterized by the chemical potential $\mu$, the eigenvalue associated with the Hamiltonian in Eq.~\eqref{eq:gpe}. 
We present length, time and energy in units of $l_x=\sqrt{\hbar/m \omega_x}$, $\omega_x^{-1}$ and $\hbar \omega_x$, respectively. We quantify the atomic interactions by the dimensionless parameter $g=mNg_0/\sqrt{2\pi}\hbar^2 l_z$. After obtaining the vortex-free condensate solution [time-independent solution to Eq.~\eqref{eq:gpe}], a vortex is imposed and the subsequent dynamics  simulated by numerical integration of the GP equation (further details in the appendix).

\begin{figure}
\includegraphics[width=0.80\textwidth]{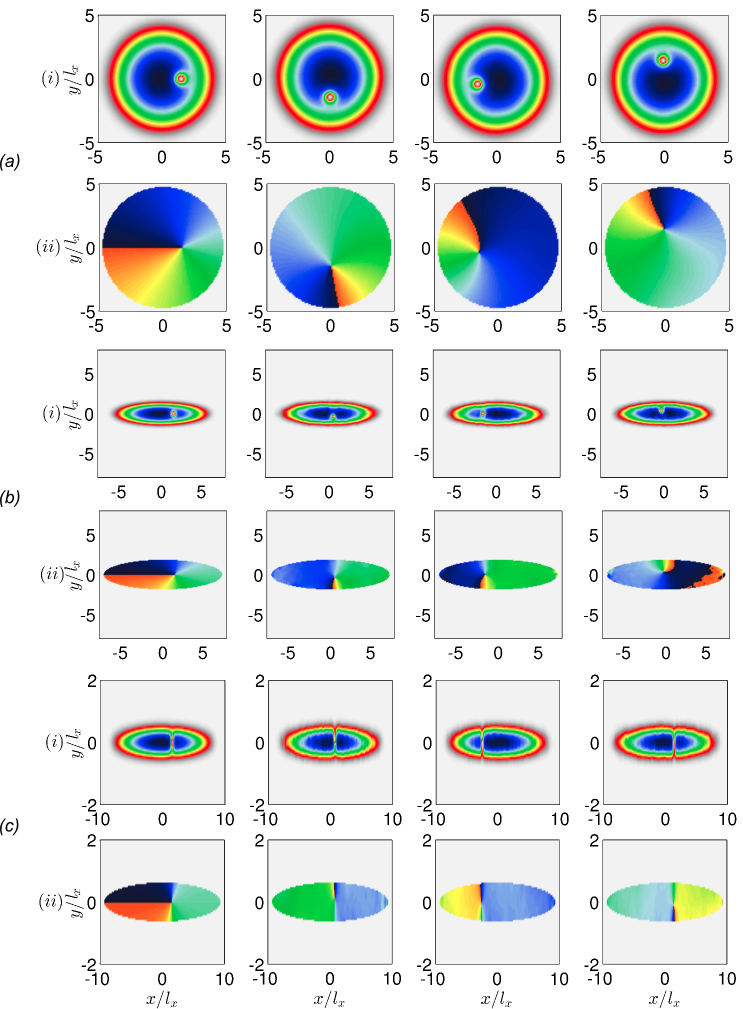}
\includegraphics[width=0.25\textwidth]{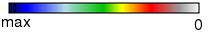} \hspace{-12mm}
\caption{(color online) Evolution of the condensate (i) density and (ii) phase profiles for trap ratios (a) $\omega_y/\omega_x=1$, (b) $4$ and (c) $15$. We take $g=400$, and the vortex initial position $(x_{V,0},y_{V,0})=(1.5l_x,0)$. 
From left to right the columns represent $t=0$, $T_{V}/4$, $T_{V}/2$ and $3T_{V}/4$, where $T_{V}=2\pi/\omega_{V}$ is the vortex precession period [calculated from Eq.~\eqref{eq:omv}].}
\label{fig:allratios}
\end{figure}

{\it Evolution for various aspect ratios.} To illustrate the crossover from vortices to solitonic vortices,  Fig.~\ref{fig:allratios} shows the condensate evolution (density and phase) under three different trap ratios, with the vortex initially at $(x_{V,0},y_{V,0})=(1.5l_x,0)$. For a circular trap ($\omega_y/\omega_x=1$) and a weakly elongated trap ($\omega_y/\omega_x=4$), the vortex precesses in a circular and elliptical path, respectively. This is to be expected since, in the absence of thermal dissipation, vortices follow equipotential trajectories  \cite{Fetter2001b}, which can be understood in terms of the Magnus force acting on the vortex due to the inhomogeneous density.  The vortex maintains a circular core and a $2\pi$ corkscrew phase profile. 
For considerably higher trap ratio ($\omega_y/\omega_x=15$), however, the initially-imprinted vortex rapidly deforms into a stripe-like density depression, and the phase profile becomes more step-like (with a rapid variation at the poles and almost uniform at the sides). As we will show, this structure behaves as a solitonic vortex.

\begin{figure*}
\centering
\includegraphics[width=1.0\textwidth]{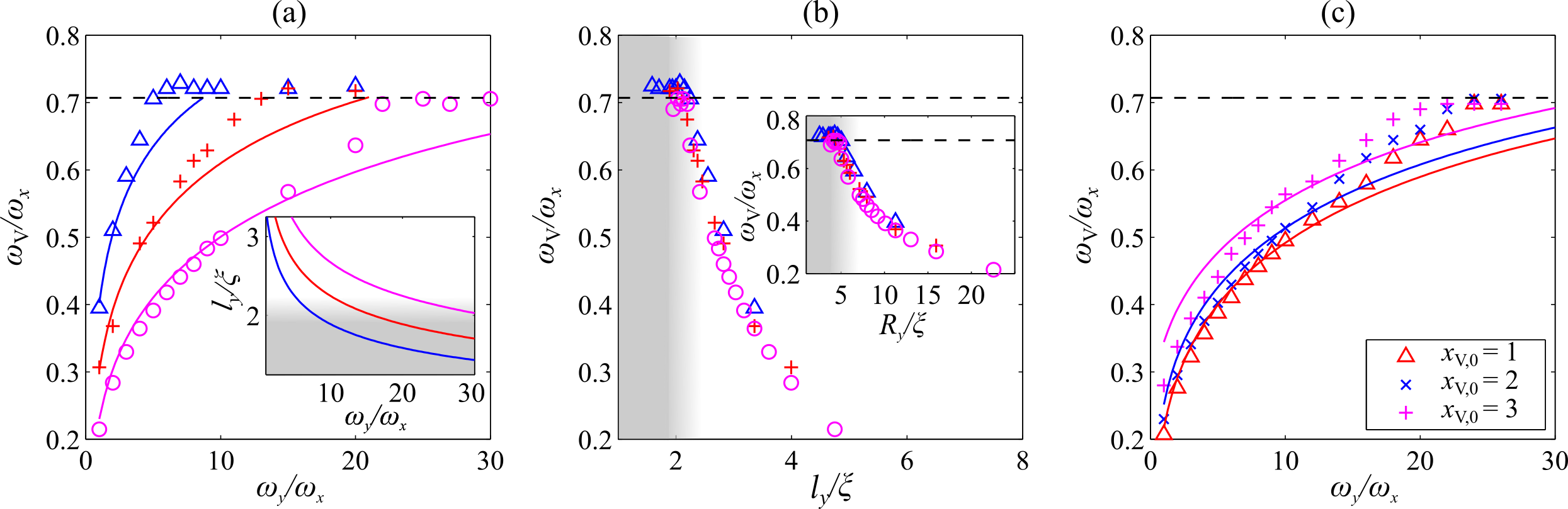}
\caption{(color online) (a) Vortex/solitonic vortex oscillation frequency $\omega_{V}$ versus trap ratio $\omega_y/\omega_x$ for interaction strengths $g=100$ (blue), $200$ (red) and $400$ (magenta), according to simulations (markers) and Eq.~\eqref{eq:omv} (lines).  The inset shows the relationship between $l_y/\xi$ and the trap ratio. The vortex is started at $(x_{V,0},y_{V,0})=(1.5l_x,0)$. Shading indicates the solitonic regime. (b) As in (a) but with $\omega_V$ plotted versus $l_{y}/\xi$, and also versus $R_y/\xi$ (inset). (c) Oscillation frequency versus trap ratio for different initial positions $x_{V,0}$, as per the legend (and $g=400$). The black dashed lines indicate the predicted soliton frequency $\omega_S=\omega_x/\sqrt{2}$. }
\label{fig:omega}
\end{figure*}

To assess how vortex-like or soliton-like the excitation is, we monitor its oscillation frequency, $\omega_V$. The excitation's trajectory $\{x_V(t),y_V(t)\}$ is tracked according to its density minimum \cite{note_on_position}; $\omega_V$ is determined from the Fourier frequency spectrum of  $x_V(t)$. A single vortex is predicted to precess with a relatively small frequency which depends non-trivially on its position, the trap frequencies and the atomic interactions \cite{Fetter2001}, as predicted using asymptotic expansions  \cite{Svidzinsky2000a,Svidzinsky2000b} and variational techniques \cite{Lundh2000,Kim2004}.  Meanwhile, for a 1D condensate in the Thomas-Fermi (TF) limit ($Na_{\rm s}/l_x \gg 1$), a dark soliton is expected to oscillate at a frequency $\omega_S=\omega_x/\sqrt{2}$ \cite{Fedichev1999,Muryshev1999,Busch2000,Huang2002}.  Figure~\ref{fig:omega}(a) plots $\omega_V$ measured across multiple simulations with varying trap ratio.  For each interaction strength $g$ considered, $\omega_V$ is seen to increase with trap ratio $\omega_y/\omega_x$, saturating at a value close to the expected dark soliton frequency $\omega_S=\omega_x/ \sqrt{2}$ (blue dashed line).  This demonstrates the soliton-like behavior of the excitation for sufficiently high trap ratios. Note that $\omega_V$ does not exactly tend to $\omega_x/\sqrt{2}$; this prediction assumes the one-dimensional and TF limits. Away from these limits the soliton frequency can deviate by up to 10\% \cite{Theocharis2007}, consistent with our observations.

It is evident from Fig.~\ref{fig:omega}(a) that, for higher $g$ values, the solitonic limit requires higher trap ratios. It is expected that the solitonic (quasi-1D) limit is reached when the transverse size of the system becomes of the order of the healing length, $\xi=\hbar/\sqrt{2mg}$, which characterizes the size of the vortex \cite{Brand2002}. Since the healing length scales as $1/\sqrt{g}$, for larger values of $g$ tighter confinement in $y$ is required to reach this limit. We formalize this criterion as follows.
In the TF approximation, density gradients are neglected and the resulting density from Eq.~\eqref{eq:gpe} takes the form $n(x,y)=n_0(1-x^2/R_x^2-y^2/R_y^2)$, where $n_0$ is the central (2D) density and $R_{x,y}^{2}=2\mu/m\omega_{x,y}^2$ define the TF radii in the respective directions. 
Applying the normalization $N=\int n(x,y){\rm d}x{\rm d}y$ leads to
\begin{equation}
\mu=\hbar\omega_{x}\sqrt{2\sqrt{\frac{2}{\pi}}\frac{a_sN}{l_z}\frac{\omega_y}{\omega_x}}.
\label{eq:mu}
\end{equation} 
Then, the ratio of the transverse harmonic oscillator length $l_y=\sqrt{\hbar/m \omega_y}$ (which characterises the transverse condensate width) to the healing length $\xi$ follows as
\begin{equation}
\frac{l_y}{\xi}=\sqrt{\frac{2 \mu}{\hbar \omega_y}}=\left(\frac{4 g \omega_x}{\pi \omega_y}\right)^{1/4}.
\label{eq:ratio}
\end{equation}
This is plotted as a function of trap ratio in Fig.~\ref{fig:omega}(a) (inset).  Upon plotting the oscillation frequency $\omega_V$ as a function of $l_y/\xi$ (rather than trap ratio), the data fall onto a common curve [Fig.~\ref{fig:omega}(b)].  Moreover, the solitonic limit is commonly reached when $l_y/\xi \ltsimeq 2$ (shaded region). One may instead define the transverse condensate as the scaled TF radius $R_y/\xi$  [inset of Fig.~\ref{fig:omega}(b)]; again the data fall onto a common curve, with the solitonic limit reached for $R_y/\xi  \ltsimeq 5$ (shaded region).

In the 2D regime ($l_y/\xi\gg 2$) the excitation should behave as a true 2D vortex. Using a variational Lagrangian method in the TF limit, Kim and Fetter \cite{Fetter2001} predicted elliptical vortex trajectories under non-axisymmetric 2D harmonic confinement, governed by the equations $\dot{x}_{V}(t)=-(\omega_{y}/\omega_{x})\omega_{V}y_{V}(t)$ and $\dot{y}_{V}(t)=(\omega_{x}/\omega_{y})\omega_{V}x_{V}(t)$, where the precession frequency $\omega_{V}$ of the vortex is defined as
\begin{equation}
\omega_{\rm V}=\frac{3}{2}\frac{\hbar}{mR_xR_y}\ln\left(\frac{R_\perp}{\xi} \right)\frac{1}{1-r_0^2},
\label{eq:omv}
\end{equation}
with $R_\perp^2=2R_x^2R_y^2/(R_x^2+R_y^2)$ and $r_0$ the radial coordinate of the vortex scaled in TF units, i.e. $r_{0}^{2}=(x_{V,0}/R_x)^{2}+(y_{V,0}/R_y)^{2}$.
This agrees well with vortex precession frequencies measured experimentally \cite{Serafini2015} for a vortex line in an elongated 3D condensate. It also  agrees well with the present simulated vortex dynamics up to moderate trap ratios, beyond which $\omega_V$ is underestimated [solid lines, Fig~\ref{fig:omega}(a)]. This difference is likely due to both the deviation from a TF state as the trap ratio increases, and also the breakdown of the assumption of a vortex phase profile used for the variational ansatz that underlies Eq.~\eqref{eq:omv}.

Figure~\ref{fig:omega}(c) shows the oscillation frequency for different initial vortex positions, $x_{V,0}=\{1, 2, 3 \}l_x$, at fixed interaction strength, $g=400$ (for comparison, $R_x\approx 5 l_{x}$).  The data have a similar behavior for all three positions, with the curves shifting up slightly compared to the prediction of Eq.~\eqref{eq:omv} for increasing $x_{V,0}$. The solitonic limit is reached at a similar trap ratio, $\omega_y/\omega_x \approx 20$. Good agreement with the Eq.~\eqref{eq:omv} is found for vortices placed close to the trap center, but worsens for vortices placed off-center. This is to be expected since off-center vortices probe more of the non-TF tails of the condensate.  
Importantly, the insensitivity of the solitonic limit to the vortex position serves to emphasize the primary role of the condensate aspect ratio (quantified via $l_y/\xi$ or $R_y/\xi$ in this work) in controlling the effective dimensionality of the excitation.

\begin{figure}[t]
\centering
\includegraphics[width=0.65\textwidth]{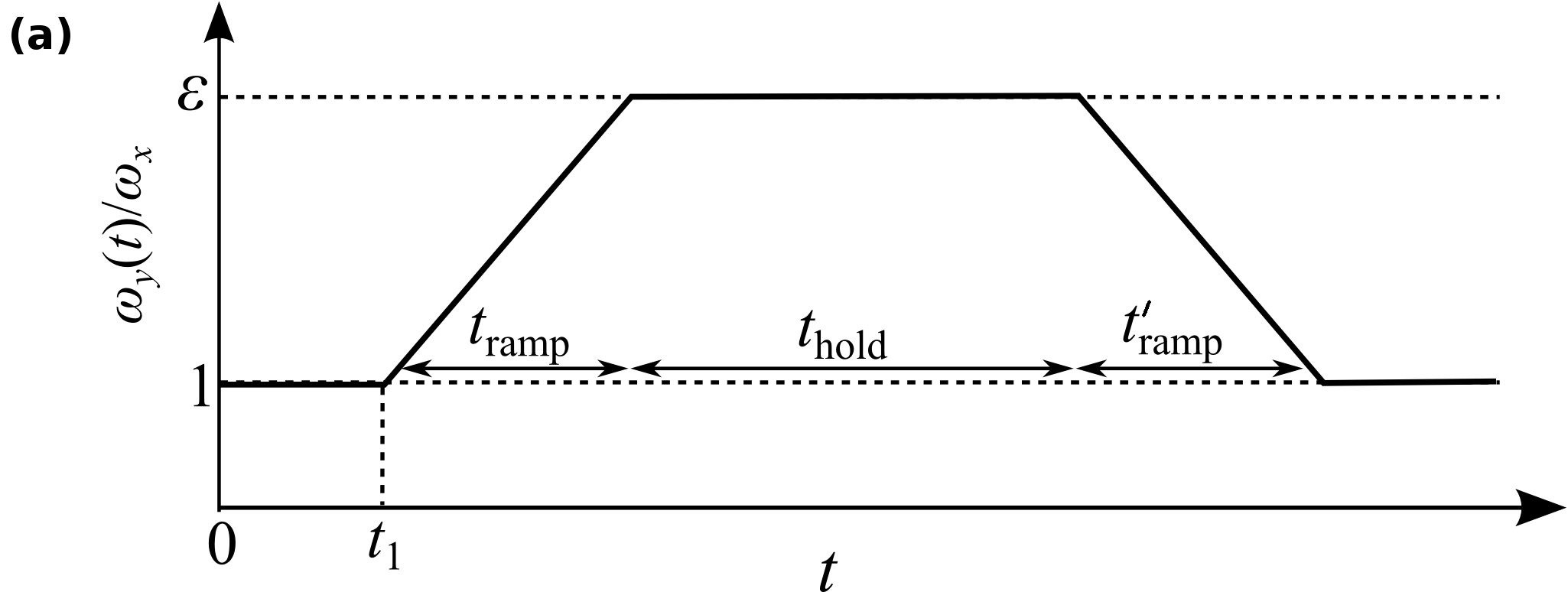}
\includegraphics[width=0.80\textwidth]{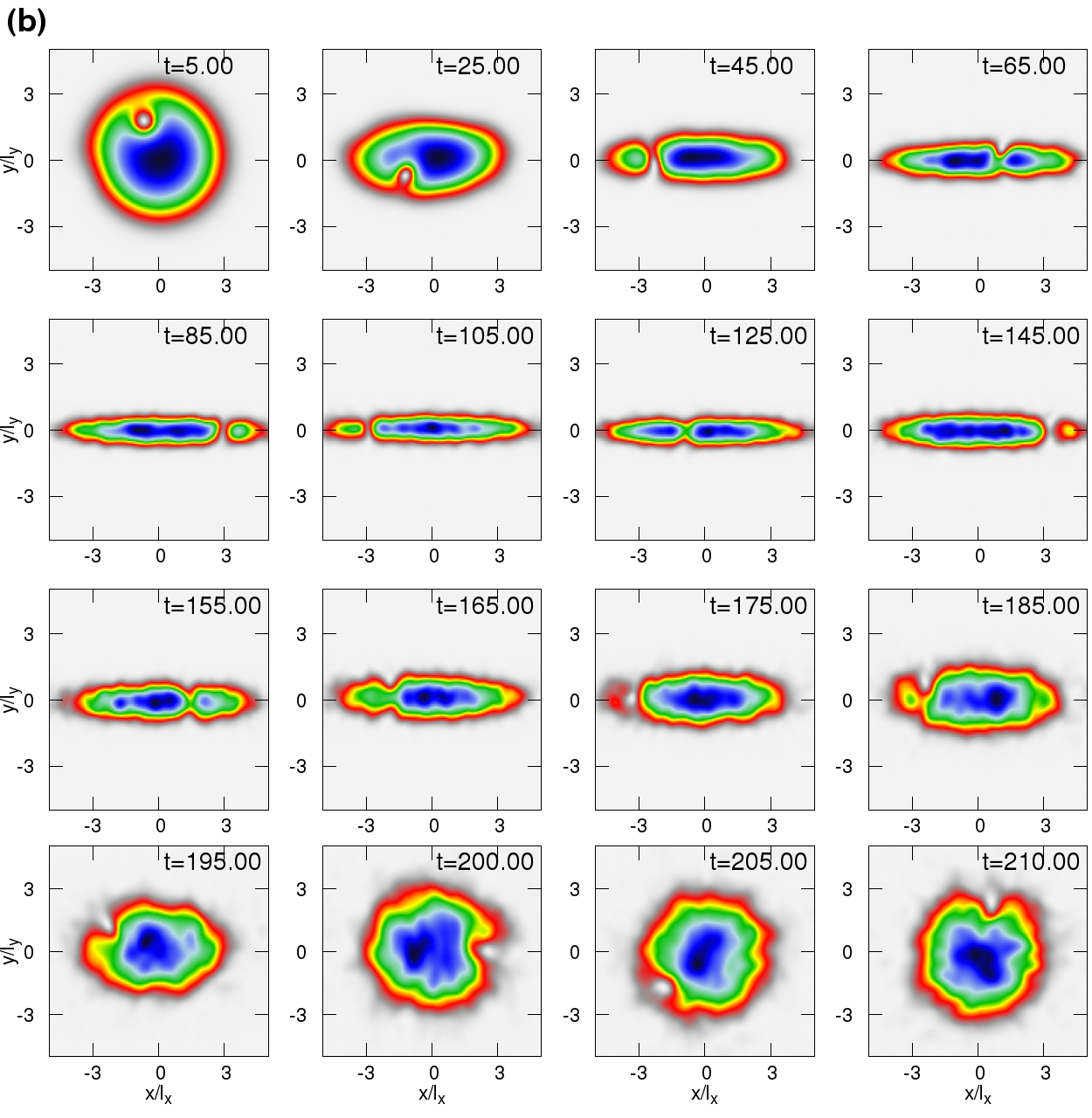}
\includegraphics[width=0.25\textwidth]{ColorBox-hori.png} \hspace{-12mm}
\caption{(color online) Dynamics under the trap deformation: (a) The imposed time-dependent deformation of the trap ratio. (b) Evolution of the condensate density during a hysteresis protocol from axisymmetric to elongated and back again. Here $g=100$, $(x_{V,0},y_{V,0})=(1.5,0)l_x$, $\varepsilon=8$, $t_{\rm{ramp}}=70\omega_x^{-1}$ and $t_{\rm{hold}}=40\omega_x^{-1}$.  
}
\label{fig:MovingRatio}
\end{figure}

{\it Evolution under trap deformation.} We now turn our attention to the fate of the vortex in a trap that is dynamically deformed from an initially axisymmetric geometry to a highly elongated (along $x$) one and back again, seeking to address the persistence of the vortex and the hysteresis of the system.  $\omega_y$ is made time-dependent so as to evolve the trap ratio $\omega_y(t)/\omega_x$ as per Fig.~\ref{fig:MovingRatio}(a): after an initial wait ($16 \omega_x^{-1}$, approximately one vortex precession period for $g=100$), the trap ratio is ramped linearly to a maximum value $\varepsilon$ over time $t_{\rm ramp}$, held there for $t_{\rm hold}$, and then linearly reduced back to an axisymmetric trap over $t_{\rm ramp}^\prime$.

\begin{figure}[t]
\includegraphics[width=0.94\textwidth]{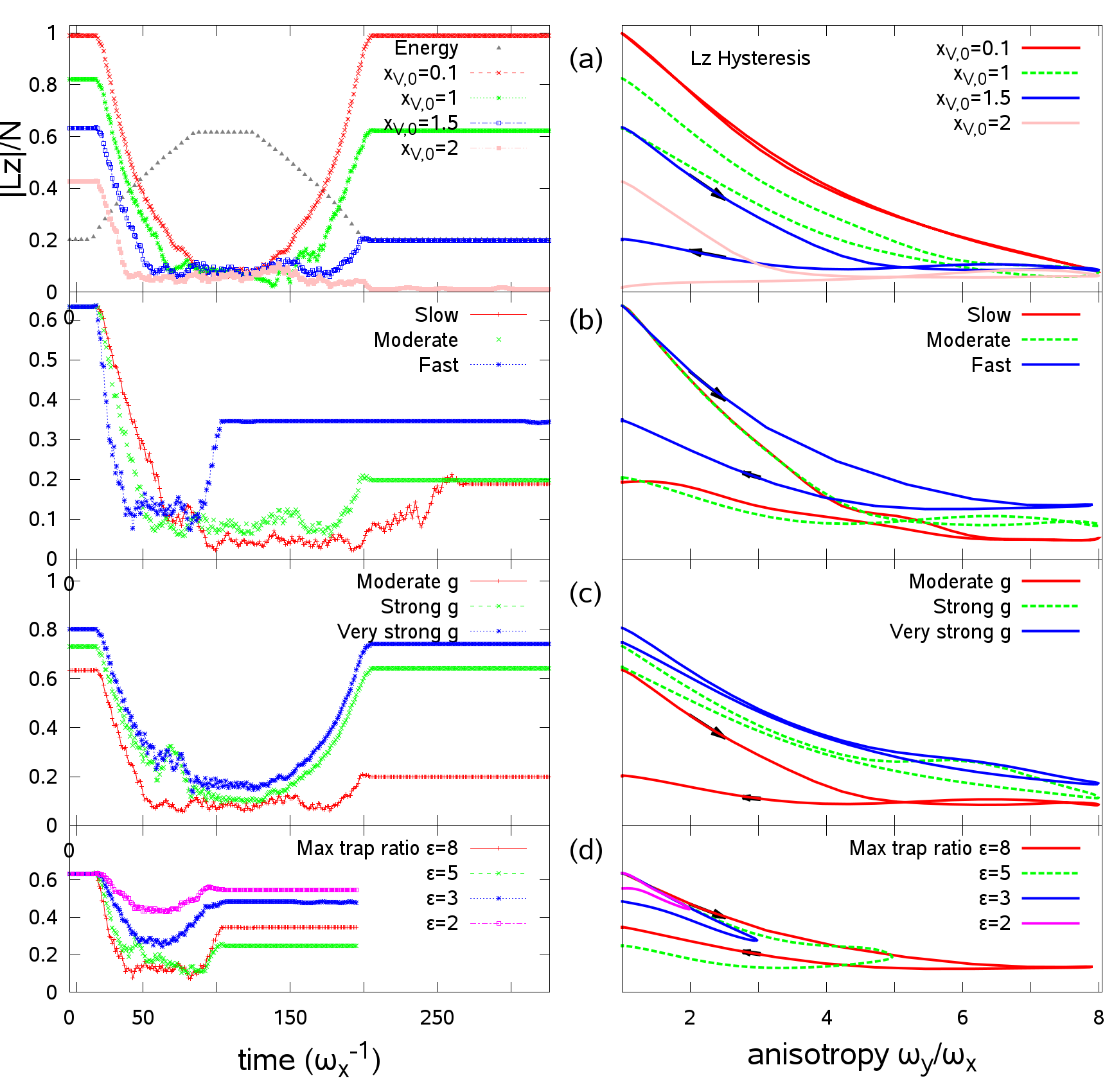}
{\caption{(color online) Angular momentum per particle, $L_z$, versus time (left) and versus the trap ratio $\omega_y/\omega_x$ (right).  From (a) to (d) we show different initial vortex positions, deformation rates, interaction strengths and maximum trap anisotropy. For the $L_z(\omega_y/\omega_x)$ hysteresis curves, the data are smoothed with Bezier curves. Unless varied, the parameter values are $g=100$, $t_{\rm{ramp}}=70\omega_x^{-1}$, $x_{V,0}=1.5l_x$ and $\varepsilon=8$. The gray line (triangle-dotted, here out of scale) of the top left panel shows the relative increase of the total energy of the system $E/E(t=0)$ that grows as large as 3. See text for details. 
}
\label{fig:AM}}
\end{figure}

An example case, with maximum trap ratio $\varepsilon=8$ and $g=100$, is shown in Fig.~\ref{fig:MovingRatio}(b). By comparison to Fig. 2(a) it is evident that, for this maximum trap ratio, the system enters the solitonic regime. It is useful to characterize the system through its total angular momentum, found from the expectation value $L_z = -i\hbar \langle \psi | x\partial_y-y\partial_x| \psi \rangle$ of the $z$-component angular momentum operator. The evolution of $L_z$ for this system is shown in Fig.~\ref{fig:AM}(a) (left column, pink line). As the trap ratio is increased, the precessing vortex deforms into a 1D-like solitonic vortex, which oscillates axially. During the increase of the trap ratio $L_z$ decreases. Upon reducing the trap ratio, the vortex is remarkably seen to re-emerge in the system, albeit with increased radial position. Concurrent with this, the angular momentum rises again, saturating at a value which is about one third of its initial value, consistent with the drift of the vortex to the edge.  The time-dependent anisotropy of the system couples with the system's nonzero angular momentum $L_z$ resulting in a smaller value of $\Lz$ than the initial one.  The condensate also develops considerable surface excitations during the deformation process.  

The gray dotted line is the variation in the total energy of the system (energy at time $t$ divided by the energy at equilibrium), here out of scale. In all cases studied, the total energy increased up to a value that follows $\sim\sqrt\varepsilon$ and then returned to a value, marginally ($\sim 0.2-2\%$) higher than the initial one, making thus the whole process a non-violent one. Since the vortex energy accounts only for $8-10\%$ of the total energy (depending on the vortex position) this comes to no surprise. We conclude that the hysteresis loop affects the angular momentum but not (significantly) the energy. 

For an adiabatically-slow deformation of the trap, $L_z$ should depend on the instantaneous trap ratio only.   It is clear here, however, that angular momentum is lost from the vortex during the dynamics, giving rise to a hysteresis effect.  This is revealed by plotting a hysteresis curve of $L_z$ versus trap ratio in Fig.~\ref{fig:AM}(a) (right column, pink line); arrows denote the direction of time.

We next make a systematic study of how the key physical parameters - vortex initial position, ramping rate, interactions and maximum trap ratio - affect the final state of the vortex and its hysteresis. The results are shown in Fig. 4(a)-(d). We vary each of these quantities in turn, while keeping the remaining parameters fixed (with values stated in the figure caption).

\begin{itemize}[leftmargin=*]
\item[] {\it (a) Vortex position}: We consider four initial vortex positions, $x_{V,0}=\{0.1,1,1.5,2\}l_x$ and $y_{V,0}=0$. For comparison, the (axisymmetric) TF radius is $R_{x}=3.4 l_x$. The loss in angular momentum increases for a vortex positioned away from the trap center.  Indeed, for positions $ \gtsimeq 0.6~R_{x}$, the remaining angular momentum is negligible, and no vortex persists. Conversely, for a vortex initially placed close to the trap center, the vortex remains intact, the system recovers its original angular momentum, and undergoes an almost time-symmetric hysteresis.

\item[] {\it (b) Deformation rate}: We consider a slow deformation ($t_{\rm ramp}=105 \omega_x^{-1}, t_{\rm hold}=40\omega_x^{-1}$), moderate deformation ($t_{\rm ramp}=70\omega_x^{-1}, t_{\rm hold}=40\omega_x^{-1}$) and fast deformation ($t_{\rm ramp}=27\omega_x^{-1}, t_{\rm hold}=20\omega_x^{-1}$). 
Note that the deformation cycles employed here are on timescales greater than the vortex periods. 

From the same initial value, the angular momentum drops at different rates. Slower deformation appears to result in an increased loss in $L_z$. However, we cannot say whether this is due to the different ramp rates, different hold times or both.  

\item[] {\it (c) Interaction}: For moderate interactions ($g=100$), angular momentum is lost during the process, while for strong ($g=200$) and very strong interactions ($g=400$) almost no angular momentum is lost and the hysteresis curve is time-symmetric. This difference is attributed to the different dimensionalities probed - for moderate interactions this system crosses the border into the solitonic regime, while for the strong and very strong cases the system remains effectively 2D throughout.  This is seen by comparison to Fig. 2 (a).

\item[] {\it (d) Maximum trap ratio}:  Lastly, we compare different maximum trap ratios, $\varepsilon=\{2,3,5,8\}$. For given interaction strength ($g=100$) these values lie around the transition from 2D to solitonic regime [see Fig.~\ref{fig:omega}(a)], and it is not surprising that the loss in angular momentum becomes larger for larger values of $\varepsilon$, i.e. as the solitonic regime is increasingly entered.
\end{itemize}

{\it Conclusions.} We explored the fate of vortices in highly elongated traps. By monitoring the precession frequency of the excitation, we mapped out the transition from vortex to solitonic-vortex, as a function of the key system parameters (trap anisotropy, interaction strength and vortex position). The frequency increases with the anisotropy and approaches the value $\omega_{\rm S}\approx \omega_{x}/\sqrt 2$, characteristic of the dark soliton oscillation.

Depending on the ratio of the healing length to the oscillator length and the initial position (initial angular momentum), the solitonic vortex will survive a continuous deformation of the trap and reappear as a vortex once the symmetry of the trap is restored (see Fig.~\ref{fig:MovingRatio}), although significant angular momentum can be lost if the solitonic regime is entered.  

Deforming and re-symmetrizing the trap that contains a solitonic vortex can be used to probe physics in scales smaller than the healing length, currently considered inaccessible to experimentalists, and could assist in current research in quantum turbulence \cite{Tsatsos2016} where the participation of several length scales is required.

Last, we mention that beyond mean-field descriptions have recently revealed how, in particular cases, quantized vorticity concurs with non trivial correlations and loss of coherence \cite{Weiner2014,Tsatsos2015,Sakmann2016}. It would be interesting to extend the present studies to fragmented condensates as well.

{\it Acknowledgments.} M.C.T acknowledges financial support from FAPESP and  CePOF - Centro de Pesquisa em \'{O}ptica e Fot\^onica and computational time on Hornet/HazelHen facilities of the High Performance Computing Center Stuttgart(HLRS). Axel Lode is thanked for having developed numerical routines associated with the calculation of angular momentum. 
We also thank P. G. Kevrekidis for useful comments. 
M.J.E and N.G.P. acknowledge support by EPSRC (UK) Grant No. EP/M005127/1.

{\it Appendix.} For the static trap simulations, the GP equation is evolved numerically using a Crank-Nicolson scheme on a spatial grid with typical spacing $\Delta x=0.05l_x$. The initial state [featuring a vortex at $(x_{V,0},y_{V,0})$] is found by imaginary time propagation of the GP equation \cite{Minguzzi2004} while enforcing a vortex phase defect $\phi_{\rm V}(x,y)=\arctan[(y-y_{V,0})/(x-x_{V,0})]$. Meanwhile, for the time-dependent simulations, the initial state is defined as $\psi_{\text{back}} A(x,y) e^{i \phi_{\rm V}(x,y)}$, with $A = \sqrt{\frac{X^2+Y^2}{X^2+Y^2+\delta}}$, $X=(x-x_{V,0})/\sigma_x,~Y= (y-y_{V,0})/\sigma_y$ and $\psi_{\text{back}}$ the vortex-free background state (found by imaginary-time propagation). 
The parameters $\delta,\sigma_x,\sigma_y$, which determine the shape of the vortex, are determined by energy minimization. The system is evolved using the MCTDH-X package \cite{ultracold}, taking $N=100$ and $M=1$.

\bibliographystyle{apsrev}

\begin{thebibliography}{99}
%
\bibitem{Bloch2008}
I. Bloch, J. Dalibard, and W. Zwerger, {\it Rev. Mod. Phys.} {\bf 80}, 885 (2008).

\bibitem{Fetter2009} A. L. Fetter, {\it Rev. Mod. Phys.} {\bf 81}, 647 (2009).

\bibitem{Anderson2010} B. P. Anderson, {\it J. Low Temp. Phys.} {\bf 161}, 574 (2010).

\bibitem{Frantzeskakis2010}
D. J. Frantzeskakis, {\it J. Phys. A} {\bf 43}, 213001 (2010).

\bibitem{Matthews1999} M. R. Matthews, B. P. Anderson, P. C. Haljan, D. S. Hall, C. E. Wieman and E. A. Cornell, {\it Phys. Rev. Lett.} \textbf{83}, 2498  (1999).

\bibitem{Rosenbusch2002} P. Rosenbusch, V. Bretin and J. Dalibard, {\it Phys. Rev. Lett.} {\bf 89}, 200403 (2002).

\bibitem{Madison2000} K. W. Madison, F. Chevy, W. Wohlleben and J. Dalibard, {\it Phys. Rev. Lett.} {\bf 84}, 806 (2000).

\bibitem{Aboshaeer2001} J. R. Abo-Shaeer, C. Raman, J. M. Vogels and W. Ketterle, {\it Science} {\bf 292}, 5516 (2001).

\bibitem{Hodby2001} E. Hodby, G. Hechenblaikner, S. A. Hopkins, O. M. Marago and C. J. Foot, {\it Phys. Rev. Lett.} {\bf 88}, 010405 (2001).

\bibitem{Anderson2001} B. P. Anderson, P. C. Haljan, C. A. Regal, D. L. Feder, L. A. Collins, C. W. Clark, and E. A. Cornell, {\it Phys. Rev. Lett.} {\bf 86}, 2926 (2001).

\bibitem{Neely2010} T.~W. Neely, E.~C. Samson, A.~S. Bradley, M.~J. Davis and B.~P. Anderson, {\it Phys. Rev. Lett.} {\bf 104}, 160401 (2010).

\bibitem{Kwon2015} W. J. Kwon, S. W. Seo and Y.-i. Shin, Phys. Rev. A {\bf 92}, 033613 (2015).

\bibitem{Freilich2010} D. V. Freilich, D. M. Bianchi, A. M. Kaufman, T. K. Langin and D. S. Hall, {\it Science} {\bf 3}, 1182-1185 (2010).

\bibitem{Henn2009} E. A. L. Henn, J. A. Seman, G. Roati, K. M. F. Magalh\~{a}es and V. S. Bagnato, {\it Phys. Rev. Lett.} {\bf 103}, 045301
(2009).

\bibitem{Neely2013} T. W. Neely, A. S. Bradley, E. C. Samson, S. J. Rooney, E. M. Wright, K. J. H. Law, R. Carretero-Gonz{\'a}lez, P. G. Kevrekidis, M. J. Davis, and B. P. Anderson,  {\it Phys. Rev. Lett.} {\bf 111}, 235301 (2013).

\bibitem{Kwon2014} W. J. Kwon, G. Moon, J. Y. Choi, S. W. Seo and Y. I. Shin, {\it Phys. Rev. A} {\bf 90}, 063627 (2014).

\bibitem{Reinhardt1997} W. P. Reinhardy and C. W. Clark, {\it J. Phys. B} {\bf 30}, L785 (1997).

\bibitem{Burger1999}
S. Burger, K. Bongs, S. Dettmer, W. Ertmer, K. Sengstock, A. Sanpera, G. V. Shlyapnikov, and M. Lewenstein, {\it Phys. Rev. Lett.} {\bf 83}, 5198 (1999).

\bibitem{Denschlag2000}
J. Denschlag, J. E. Simsarian, D. L. Feder, C. W. Clark, L. A. Collins, J. Cubizolles, L. Deng, E. W. Hagley, K. Helmerson, W. P. Reinhardt, S. L. Rolston, B. I. Schneider, W. D. Phillips, {\it Science} {\bf 287}, 97 (2000).

\bibitem{Dutton2001} Z. Dutton, M. Budde, C. Slowe and L. V. Hau, {\it Science} {\bf 293}, 663 (2001).

\bibitem{Jo2007} G.-B. Jo, J.-H. Choi, C. A. Christensen, T. A. Pasquini, Y.-R. Lee, W. Ketterle, and D. E. Pritchard, {\it Phys. Rev. Lett.} {\bf 98}, 180401 (2007).

\bibitem{Engels2007} P. Engels and C. Atherton, {\it Phys. Rev. Lett.} {\bf 99}, 160405 (2007).

\bibitem{Becker2008}
C. Becker, S. Stellmer, P. S.-Panahi, S. D\"orscher, M. Baumert, E.-M. Richter, J. Kronj\"ager, K. Bongs, and K. Sengstock, {\it Nat. Phys.} {\bf 4}, 496 (2008). 

\bibitem{Chang2008} J. J. Chang, P. Engels and M. A. Hoefer, {\it Phys. Rev. Lett.} {\bf 101}, 170404 (2008).

\bibitem{Stellmer2008}
S. Stellmer, C. Becker, P. Soltan-Panahi, E.-M. Richter, S. D\"orscher, M. Baumert, J. Kronj\"ager, K. Bongs, and K. Sengstock, {\it Phys. Rev. Lett.} {\bf 101}, 120406 (2008).

\bibitem{Weller2008}
A. Weller, J. P. Ronzheimer, C. Gross, J. Esteve, M. K. Oberthaler, D. J. Frantzeskakis, G. Theocharis, and P. G. Kevrekidis, {\it Phys. Rev. Lett.} {\bf 101}, 130401 (2008).

\bibitem{Fedichev1999} P. O. Fedichev, A. E. Muryshev, and G. V. Shlyapnikov, {\it Phys. Rev. A} {\bf 60}, 3220 (1999).

\bibitem{Muryshev1999} A. E. Muryshev, H. B. van Linden van den Heuvell, and G. V. Shlyapnikov, {\it Phys. Rev. A} {\bf 60}, R2665 (1999).

\bibitem{Busch2000} T. Busch and J. R. Anglin, {\it Phys. Rev. Lett.} {\bf 84}, 2298 (2000).

\bibitem{Huang2002} G. Huang, J. Szeftel and S. Zhu, {\it Phys. Rev. A} {\bf 65}, 053605 (2002).

\bibitem{Svidzinsky2000a} A. A. Svidzinsky and A. L. Fetter, {\it Phys. Rev. Lett.} {\bf 84}, 5919 (2000).

\bibitem{Svidzinsky2000b} A. A. Svidzinsky and A. L. Fetter, {\it Phys. Rev. A} {\bf 62}, 063617 (2000).

\bibitem{Lundh2000} E. Lundh and P. Ao, {\it Phys. Rev. A} {\bf 61}, 063612 (2000).

\bibitem{Kim2004} J.-K. Kim and A. L. Fetter, {\it Phys. Rev. A} {\bf 70}, 043624 (2004).

\bibitem{Middelkamp2010} S. Middelkamp,  P. G. Kevrekidis, D. J. Frantzeskakis, R. Carretero-Gonz{\'a}lez, and P. Schmelcher, {\it Phys. Rev. A} {\bf 82}, 013646 (2010).

\bibitem{Lamporesi2013} G. Lamporesi, S. Donadello, S. Serafini, F. Dalfovo and C. Ferrari, {\it Nat. Phys.} {\bf 9}, 656 (2013).

\bibitem{Weiler2008} C. N. Weiler, T. W. Neely, D. R. Scherer, A. S. Bradley, M. J. Davis and B. P. Anderson, {\it Nature} {\bf 455}, 948 (2008).

\bibitem{Middelkamp2011} S. Middelkamp, P. J. Torres, P. G. Kevrekidis, D. J. Frantzeskakis, R. Carretero-Gonz{\'a}lez, P. Schmelcher, D. V. Freilich, D. S. and Hall, {\it Phys. Rev. A.} {\bf 84}, 011605 (2011).

\bibitem{Frisch1992}T. Frisch, Y. Pomeau, and S. Rica, {\it Phys. Rev. Lett.} {\bf 69}, 1644 (1992).

\bibitem{Inouye2001} S. Inouye, S. Gupta, T. Rosenband, A. P. Chikkatur, A. G\"orlitz, T. L. Gustavson, A. E. Leanhardt, D. E. Pritchard, and W. Ketterle, {\it Phys. Rev. Lett.} {\bf 87}, 080402 (2001).

\bibitem{Hakim1997} V. Hakim, Phys. {\it Rev. E} {\bf 55}, 2835 (1997).

\bibitem{Parker2003} N. G. Parker, N. P. Proukakis, M. Leadbeater and C. S. Adams, {\it Phys. Rev. Lett.} {\bf 90}, 220401 (2003).

\bibitem{Parker2004} N. G. Parker, N. P. Proukakis, C. F. Barenghi and C. S. Adams, {\it Phys. Rev. Lett.} {\bf 92}, 160403 (2004).

\bibitem{Allen2011}  A. J. Allen, D. P. Jackson, C. F. Barenghi and N. P. Proukakis, {\it Phys. Rev. A} {\bf 83}, 013613 (2011).

\bibitem{Parker2012} N. G. Parker, A. J. Allen, C. F. Barenghi and N. P. Proukakis, {\it Phys. Rev. A} {\bf 86}, 013631 (2012).

\bibitem{Theocharis2007} G. Theocharis, P. G. Kevrekidis, M. K. Oberthaler and D. J. Frantzeskakis, {\it Phys. Rev. A} {\bf 76}, 045601 (2007).

\bibitem{Kivshar1998} Y. S. Kivshar and B. Luther-Davies, {\it Phys. Rep.} {\bf 278}, 81 (1998).

\bibitem{Feder2000} D. L. Feder, M. S. Pindzola, L. A. Collins, B. I. Schneider and C. W. Clark, {\it Phys. Rev. A} {\bf 62}, 053606 (2000).

\bibitem{Carr2000} L. D. Carr, M. A. Leung and W. P. Reinhardy, {\it J. Phys. B: At. Mol. Opt. Phys.} {\bf 33}, 3983 (2000).

\bibitem{Ginsberg2005} N. S. Ginsberg, J. Brand and L. V. Hau, {\it Phys. Rev. Lett.} {\bf 94}, 040403 (2005).

\bibitem{Brand2002}
J. Brand and W. P. Reinhardt, {\it Phys. Rev. A} \textbf{65}, 043612, (2002).

\bibitem{Komineas2003}
S. Komineas and N. Papanicolaou, {\it Phys. Rev. A} \textbf{68}, 043617 (2003).

\bibitem{Becker2013}
C. Becker, K. Sengstock, P. Schmelcher, P. G. Kevrekidis, and R. Carretero-Gonz{\'a}lez,
{\it New J. Phys.} \textbf{15}, 113028 (2013).

\bibitem{Donadello2014}
S. Donadello, S. Serafini, M. Tylutki, L. P. Pitaevskii, F. Dalfovo, G. Lamporesi and G. Ferrari,  {\it Phys. Rev. Lett.} \textbf{113}, 065302 (2014).

\bibitem{Ku2014} 
M. J. H. Ku, W. Ji, B. Mukherjee, E. Guardado-Sanchez, L. W. Cheuk, T. Yefsah and M. W. Zwierlein, {\it Phys. Rev. Lett.} \textbf{113}, 065301 (2014).

\bibitem{Gorlitz2001}
A. G\"{o}rlitz, J. M. Vogels, A. E. Leanhardt, C. Raman, T. L. Gustavson, J. R. Abo-Shaeer, A. P. Chikkatur, S. Gupta, S. Inouye, T. Rosenband, and W. Ketterle, {\it Phys. Rev. Lett.} {\bf 87}, 130402 (2001).

\bibitem{Fetter2001b} A. L. Fetter and A. A. Svidzinsky, {\it J. Phys-Condens. Mat.} {\bf 13}, R135 (2001).

\bibitem{note_on_position} For high aspects ratios close to, or within, the solitonic vortex regime, the transverse position of the excitation is ill-determined, and so only the axial position is evaluated.

\bibitem{Fetter2001} A. L. Fetter and J. K. Kim, {\it J. Low Temp. Phys.} {\bf 125}, 239 (2001).

\bibitem{Serafini2015} S. Serafini, M. Barbiero, M. Debortoli, S. Donadello, F. Larcher, F. Dalfovo, G. Lamporesi, and G. Ferrari, {\it Phys. Rev. Lett.} {\bf 115}, 170402 (2015).

\bibitem{Tsatsos2016}
M. C. Tsatsos, P. E. S. Tavares, A. Cidrim, A. R. Fritsch, M. A. Caracanhas, F. E. A. dos Santos, C. F. Barenghi, V. S. Bagnato, {\it Phys. Rep.} {\bf 622}, 1 (2016). 

\bibitem{Weiner2014}
S. E. Weiner, M. C. Tsatsos, L. S. Cederbaum and Lode, A. U. J, arXiv: 1409.7670 (2014).
 
\bibitem{Tsatsos2015}
M. C. Tsatsos and A. U. J. Lode, {\it J. Low Temp. Phys.} {\bf 181}, 171 (2015).

\bibitem{Sakmann2016}
K. Sakmann and   M. Kasevich, {\it Nat. Phys.}, in press, doi:10.1038/nphys3631 (2016).

\bibitem{Minguzzi2004} A. Minguzzi, S. Succi, F. Toschi, M. P. Tosi and P. Vignolo, {\it Phys. Rep.} {\bf 395}, 223 (2004).

\bibitem{ultracold}
A. U. J. Lode, M. C. Tsatsos, E. Fasshauer, MCTDH-X: The time-dependent multiconfigurational Hartree for  indistinguishable particles software, \href{http://ultracold.org}{ultracold.org}.

\end{thebibliography}

\end{document}